\documentclass[twocolumn,prl,showpacs,floatfix,amsfonts,superscriptaddress, bibnotes]{revtex4} 
 
\usepackage[]{graphicx} 
\usepackage{amsbsy} 
\usepackage{float}
 
\begin{document} 
\title{Opening of the superconducting gap in the hole pockets of Ba(Fe$_{1-x}$Co$_x$)$_2$As$_2$ as seen via Angle-Resolved PhotoElectron Spectroscopy.}

\author{B.~Mansart} 
\affiliation{Laboratoire de Physique des Solides, CNRS-UMR 8502, Universit\'{e} Paris-Sud, F-91405 Orsay, France}
\altaffiliation[]{current address: Laboratory for Ultrafast Microscopy and Electron Scattering, ICMP, Ecole Polytechnique F\'{e}d\'{e}rale de Lausanne, CH-1015 Lausanne, Switzerland}
\author{E.~Papalazarou} 
\affiliation{Laboratoire de Physique des Solides, CNRS-UMR 8502, Universit\'{e} Paris-Sud, F-91405 Orsay, France} 
\author{M.~Fuglsang Jensen} 
\affiliation{Laboratoire de Physique des Solides, CNRS-UMR 8502, Universit\'{e} Paris-Sud, F-91405 Orsay, France} 
\author{V.~Brouet} 
\affiliation{Laboratoire de Physique des Solides, CNRS-UMR 8502, Universit\'{e} Paris-Sud, F-91405 Orsay, France} 
\author{L.~Petaccia} 
\affiliation{Sincrotrone Trieste S.C.p.A., Strada Statale 14 km 163.5, I-34149 Trieste, Italy} 
\author{L.~de'~Medici} 
\affiliation{Laboratoire de Physique des Solides, CNRS-UMR 8502, Universit\'{e} Paris-Sud, F-91405 Orsay, France}
\author{G.~Sangiovanni} 
\affiliation{Institut f\"{u}r Festk\"{o}rperphysik, Technische Universit\"{a}t Wien, Vienna, Austria}
\author{F.~Rullier-Albenque} 
\affiliation{Service de Physique de l'Etat Condens\'{e}, Orme des Merisiers, CEA Saclay (CNRS URA 2464), F-91195 Gif-Sur-Yvette cedex, France} 
\author{A.~Forget} 
\affiliation{Service de Physique de l'Etat Condens\'{e}, Orme des Merisiers, CEA Saclay (CNRS URA 2464), F-91195 Gif-Sur-Yvette cedex, France} 
\author{D.~Colson} 
\affiliation{Service de Physique de l'Etat Condens\'{e}, Orme des Merisiers, CEA Saclay (CNRS URA 2464), F-91195 Gif-Sur-Yvette cedex, France}
\author{M.~Marsi}
\affiliation{Laboratoire de Physique des Solides, CNRS-UMR 8502, Universit\'{e} Paris-Sud, F-91405 Orsay, France}

\date{\today} 
 
\begin{abstract} 

We present an Angle-Resolved PhotoElectron Spectroscopy study of the changes in the electronic structure of electron doped Ba(Fe$_{1-x}$Co$_x$)$_2$As$_2$ across the superconducting phase transition. By changing the polarization of the incoming light, we were able to observe the opening of the gap for the inner hole pocket $\alpha$, and to compare its behavior with the outer hole-like band $\beta$. Measurements along high symmetry directions show that the behavior of $\beta$ is consistent with an isotropic gap opening, while slight anisotropies are detected for the inner band $\alpha$. The implications of these results for the $s\pm$ symmetry of the superconducting order parameter are discussed, in relation to the nature of the different iron orbitals contributing to the electronic structure of this multiband system. 


\end{abstract} 
 
\maketitle 


After the recent discovery of iron-based superconducting compounds~\cite{Kamihara2008, Rotter2008}, several experimental and theoretical studies tried to explain the mechanisms underlying their high critical temperatures (T$_c$). To this end, the symmetry of the order parameter is one critical piece of information~\cite{Hirschfeld2011}. For instance, considerable interest was raised by early theoretical predictions~\cite{Mazin2008, Kuroki2008}, describing the possibility of a $s\pm$ symmetry with a gap following the law $\Delta=\Delta_0(cos(k_x)+ cos(k_y))$, implying a dephasing of $\pi$ between hole and electron pockets of the Fermi surface. 

Experimentally, Angle-Resolved PhotoElectron Spectroscopy (ARPES) has been widely used in this search, thanks to its capability of resolving in k-space the electronic structure of materials~\cite{Damascelli2003}. This technique has been employed to measure the amplitude of the superconducting gap (SC gap) in all the existing iron-based compound families: the 1111 (RE(O$_{1-x}$F$_x$)FePn, RE being a rare-earth and Pn a pnictogen)~\cite{Sato2008, Aiura2008}, 11 (FeSe$_{1-x}$, FeSe$_{1-x}$Te$_x$ and FeTe$_{1-x}$S$_x$)~\cite{Nakayama2010}, 111 (LiFeAs and CaFeAs)~\cite{Borisenko2010, Umezawa2012}, in the iron selenides A$_x$Fe$_2$Se$_2$ (A=Cs, K)~\cite{Zhang2011} and in the 122 compounds (doped AEFe$_2$As$_2$, AE being an alcaline earth metal).

\begin{figure}[h] 
\includegraphics[width=0.9\linewidth,clip=true]{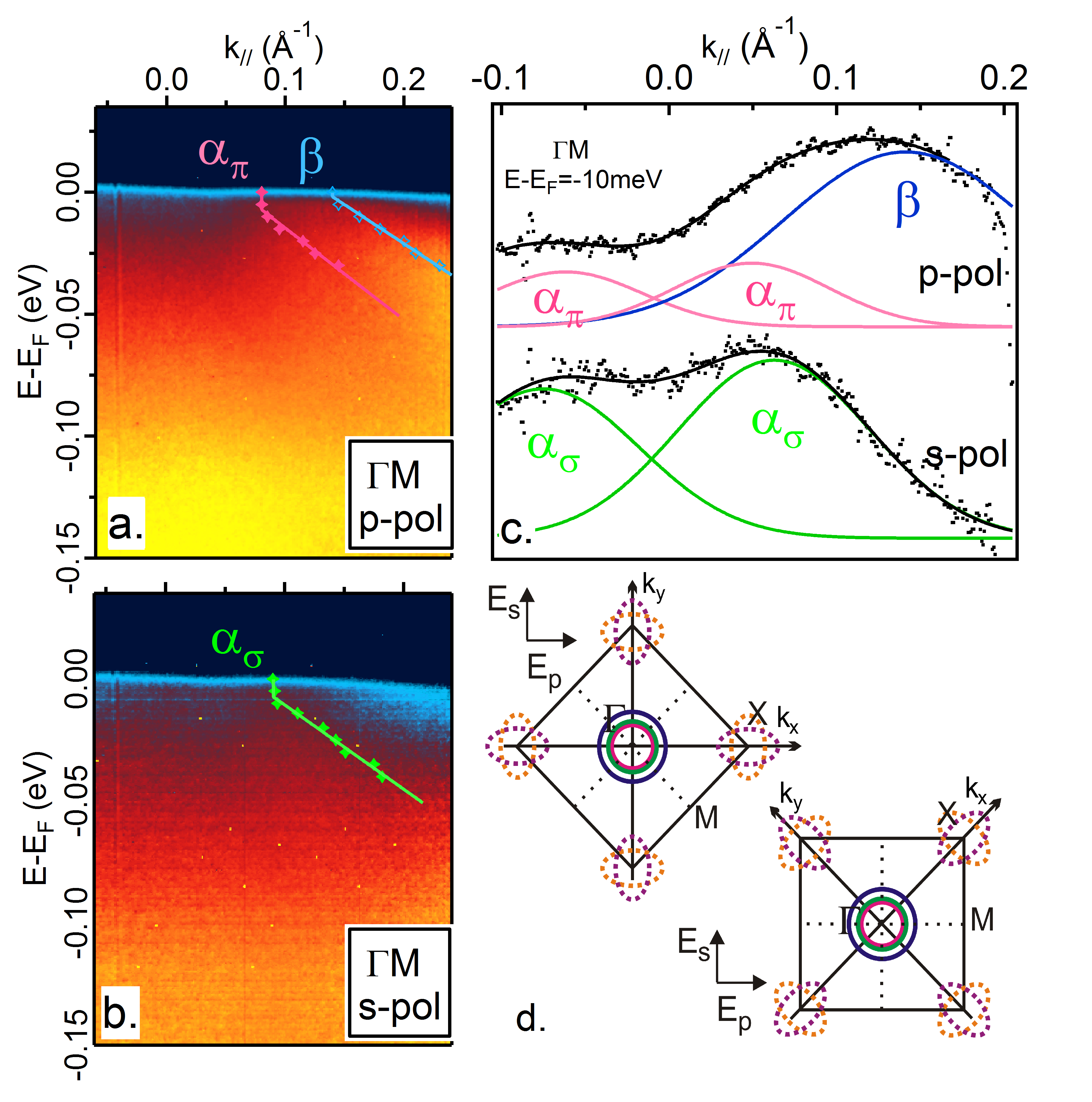} 
\caption{ARPES images acquired at T=14 K along the $\Gamma M$ sample orientation, (a) in p-polarization and (b) in s-polarization; (c) Momentum Dispersion Curves extracted from the images at E-E$_F$ = -10 meV. (d) experimental geometries for measurements along $\Gamma X$ and $\Gamma M$.} 
\label{images}
\end{figure}

In particular, this latter family has been extensively studied, due also to the availability of high quality samples. The 122 compounds present a Fermi surface composed of three hole pockets (two almost degenerate inner $\alpha$ bands and an outer $\beta$ band) located around the $\Gamma Z$ direction, and two electron pockets ($\gamma$ and $\delta$) located around $XA$. The Fermi surface and the folded first Brillouin zone in two dimensions are shown in Fig.~\ref{images} (d). In a recent study, we showed that selective measurements of these bands are possible with polarization-dependent ARPES, which also gives insight into the orbital character for each one of them~\cite{Mansart2011}. The highest T$_c$ in this family is obtained in optimally hole-doped Ba$_{0.6}$K$_{0.4}$Fe$_2$As$_2$, for which several ARPES experiments showed that the SC gap amplitude was band-dependent~\cite{Ding2008, Wray2008, Nakayama2009, Nakayama2011}, and presented a three-dimensional character for one of the hole pockets~\cite{Zhang2010, Xu2011}. However, another recent study showed an orbital independent SC gap amplitude in both hole-doped Ba$_{1-x}$K$_{x}$Fe$_2$As$_2$ and isovalent  BaFe$_{2}$(As$_{1-x}$P$_x$)$_2$~\cite{Shimojima2011}, raising the question of a possible orbital-fluctuations mediated superconductivity mechanism.

Only one ARPES measurement of the SC gap has been reported so far in the electron-doped system Ba(Fe$_{1-x}$Co$_x$)$_2$As$_2$~\cite{Terashima2009}. The SC gap was measured to be larger in $\beta$ than in $\gamma$; no information was obtained on the $\alpha$ gap, since this band could not be seen to cross the Fermi level. Interestingly, the SC gap appeared isotropic for both $\gamma$ and $\beta$.

We report here an observation of momentum and band-dependent SC gaps using polarization-dependent ARPES in Ba(Fe$_{1-x}$Co$_x$)$_2$As$_2$, which allowed us to detect the gap opening across T$_c$ for both $\alpha$ and $\beta$. These results confirm that the order parameter is consistent with a $s\pm$ symmetry. A slight anisotropy in the behavior of $\alpha$ along $\Gamma X$ and $\Gamma M$ could be detected, and we discuss how this may require the introduction of multiple harmonic terms for the description of the order parameter. 

We studied single crystal samples of Ba(Fe$_{1-x}$Co$_x$)$_2$As$_2$, x=0.07 with T$_c$=24.5~K, grown by the self-flux method~\cite{Rullier-Albenque}. They were fully characterized and oriented prior to our measurements. ARPES experiments were performed on the BaDElPh beamline at the Elettra synchrotron light source~\cite{Petaccia2009}, which is optimized for photon energies $h\nu$ below 10~eV, allowing bulk-sensitive measurements; all the results presented in this article have been obtained at a photon energy of 9 eV. The photoelectron analyser was a SPECS Phoibos 150, with an energy resolution of 5~meV, and an angular resolution of $0.1^{\circ}$. Measurements were performed on high quality surfaces cleaved in-situ under ultra-high vacuum with pressure better than $5\times10^{-11}$~mbar. The reproducibility of all the experimental results presented here was verified via multiple thermal cycles across T$_c$, yielding consistent results.


ARPES images in the $\Gamma M$ direction are presented in Fig.~\ref{images} (a) and (b). Three hole-like pockets are visible, two almost degenerate inner $\alpha$ bands (that in our previous work~\cite{Mansart2011} were named $\alpha_{\sigma}$ and $\alpha_{\pi}$) and an outer $\beta$ band, as shown in the Momentum Dispersion Curves (MDCs) in Fig.~\ref{images} (c). As shown in our previous study~\cite{Mansart2011}, photon polarization and $k_z$ dispersion effects make it possible to obtain 
information on their orbital character. Namely, even orbitals with respect to the photoemission plane are measureable with p-polarized photons, while odd ones are visible only with s-polarized photons.
Our experimental geometry is depicted in Fig.~\ref{images} (d), and the notations for sample orientation and photon polarization are the same as in Ref.~\onlinecite{Mansart2011}: in particular, $\Gamma X$ corresponds to the nearest neighbours Fe-Fe bond direction.  
In the $k_z\cong0.8~\left[4\pi/c\right]$ plane, probed with h$\nu$ = 9 eV, and along $\Gamma M$, $\beta$ has a marked $d_{z^2}$ character, which in our experimental configuration is even and thus measurable only in p-polarization. On the other hand, the $\alpha$ bands are formed by combinations of $d_{xz}$ and $d_{yz}$ orbitals, and are detectable for every sample orientation and polarization. Consequently, in p-polarization we simultaneously detect both the even $\alpha$ band and $\beta$, with $\beta$ having an overwhelming  spectral weight. Conversely, in s-polarization $\beta$'s photoemission yield is suppressed by matrix element effects, and the odd $\alpha$ band can be measured in a very accurate way, which was not possible in the previous gap study on this system~\cite{Terashima2009}. 

  \begin{figure}[h] 
\includegraphics[width=0.9\linewidth,clip=true]{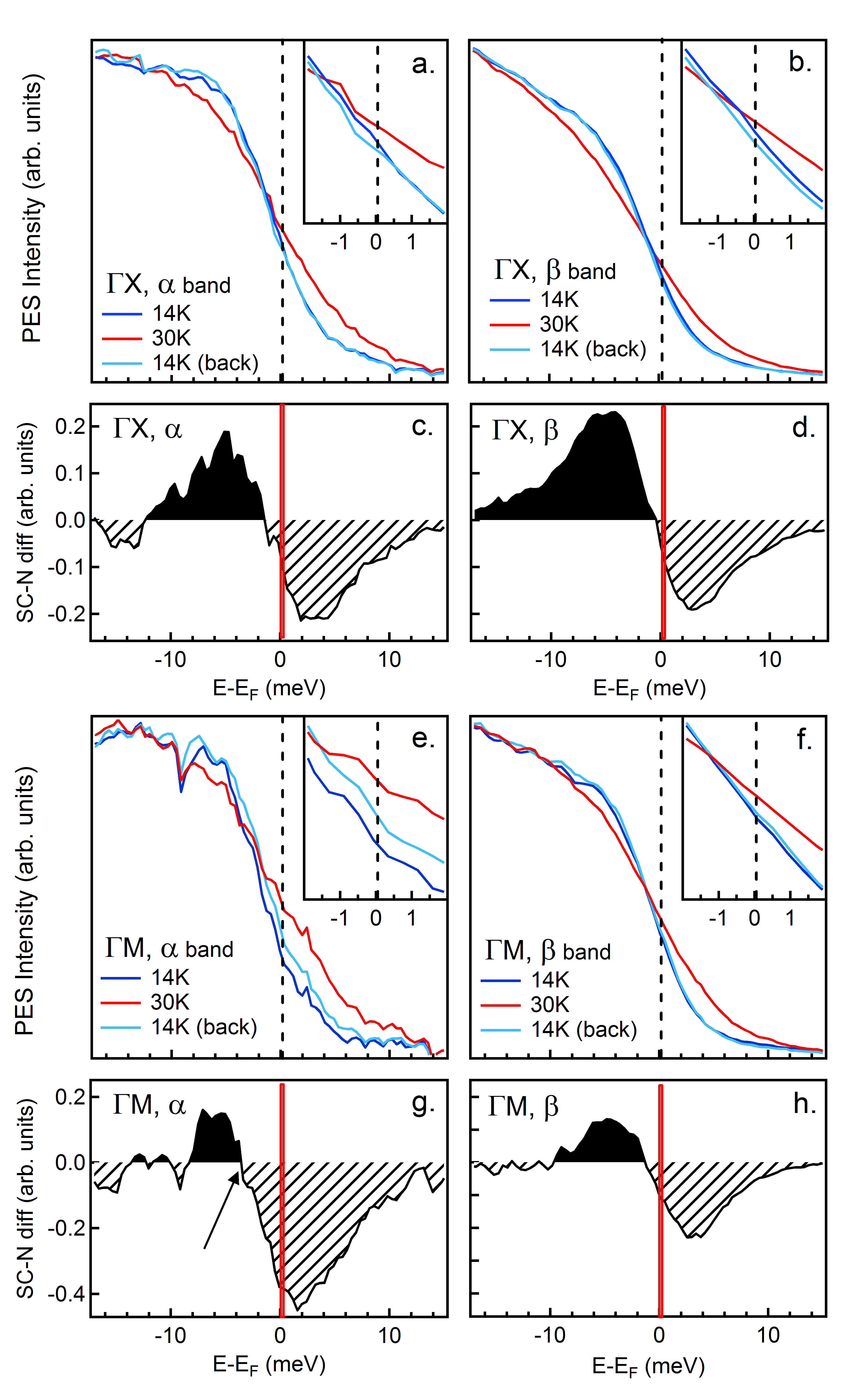} 
\caption{(a)-(c) Energy Dispersion Curves for the $\alpha$ and $\beta$-bands, along $\Gamma X$ (a-b) and $\Gamma M$ (e-f); the insets show in more detail the Fermi level region. (c-d) and (g-h) Corresponding differences between superconducting phase and normal state. Red dots in (g-h) correspond to simulations using a BCS-like function, giving for $\beta$ a coherent quasiparticle with width 3.5 meV and centered at 2 meV, and for $\alpha$ the same width but at 3.5 meV away from the Fermi level.} 
\label{gap_GM}
\end{figure}

 In Fig.~\ref{gap_GM}, we present the Energy Dispersion Curves (EDCs), extracted from ARPES images along $\Gamma M$ and $\Gamma X$, at the Fermi wavevector corresponding to each band and for two temperatures across the superconducting (SC) phase transition. Upon cooling down, a shift of the Fermi Leading Edge (FLE) is observed, corresponding to the opening of the SC gap, in every band. This shift can be also seen in the differences between the SC and normal state EDCs, shown in Fig.~\ref{gap_GM} (c), (d), (g) and (h). The simple thermal contribution to the Fermi level broadening - different at T=14 K and T=30K - would give a symmetric peak-valley feature in the differences. For our data, the zero-level crossing happens at energies lower than the Fermi energy (indicated by an arrow in Fig.~\ref{gap_GM} (g)), which is a very sensitive indication of the shift of the leading edge. 
Since no prominent quasiparticle peak could be observed in the SC state, we did not try to evaluate the gap value by modelling the spectral lineshape 
~\cite{Norman1998, Campuzano1996}. 

\begin{figure}[ht] 
\includegraphics[width=1\linewidth,clip=true]{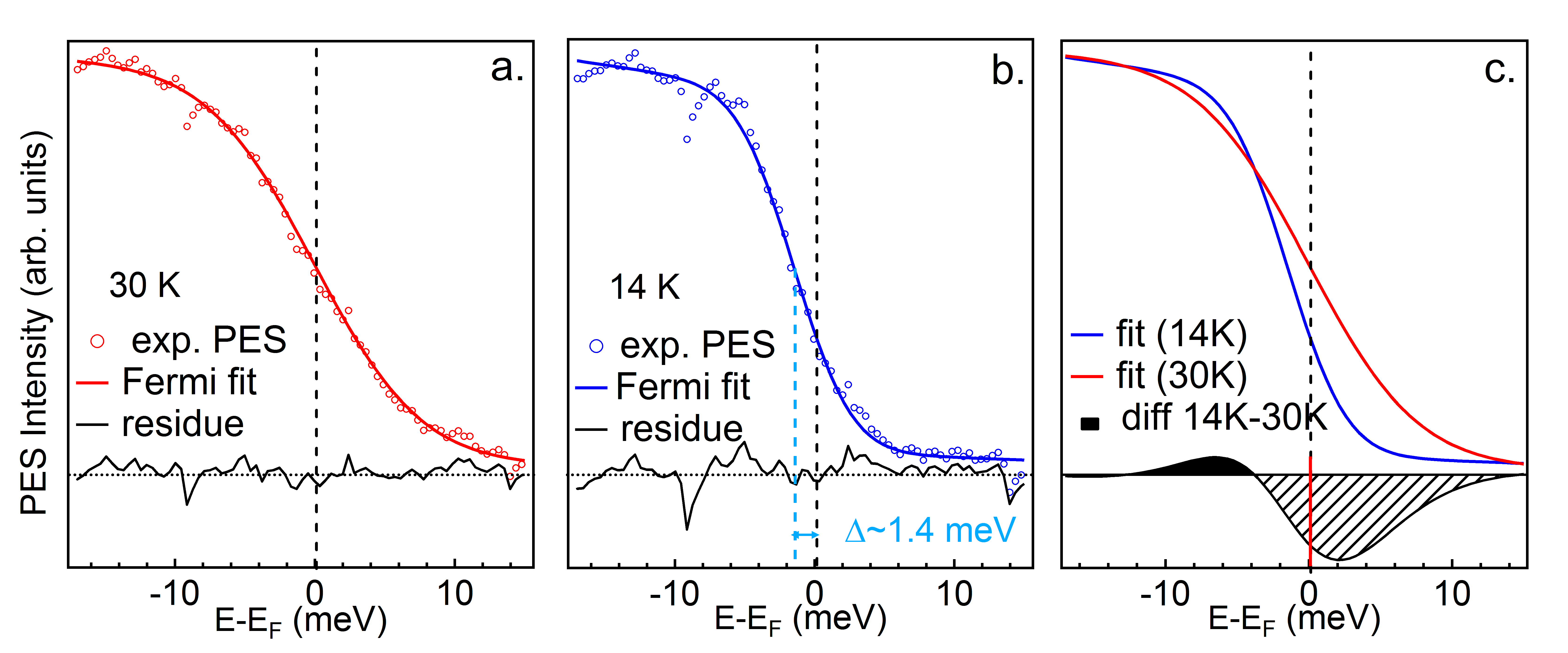} 
\caption{Energy Dispersion Curves fitting by a Fermi function, for $\alpha$ band along $\Gamma M$ at (a) 30 K and (b) 14 K. (c) Fitting functions with their difference; the Fermi level crossing is shifted towards negative energies, indicating the opening of the superconducting gap.} 
\label{fitting}
\end{figure}

\begin{figure}[h] 
\includegraphics[width=0.95\linewidth,clip=true]{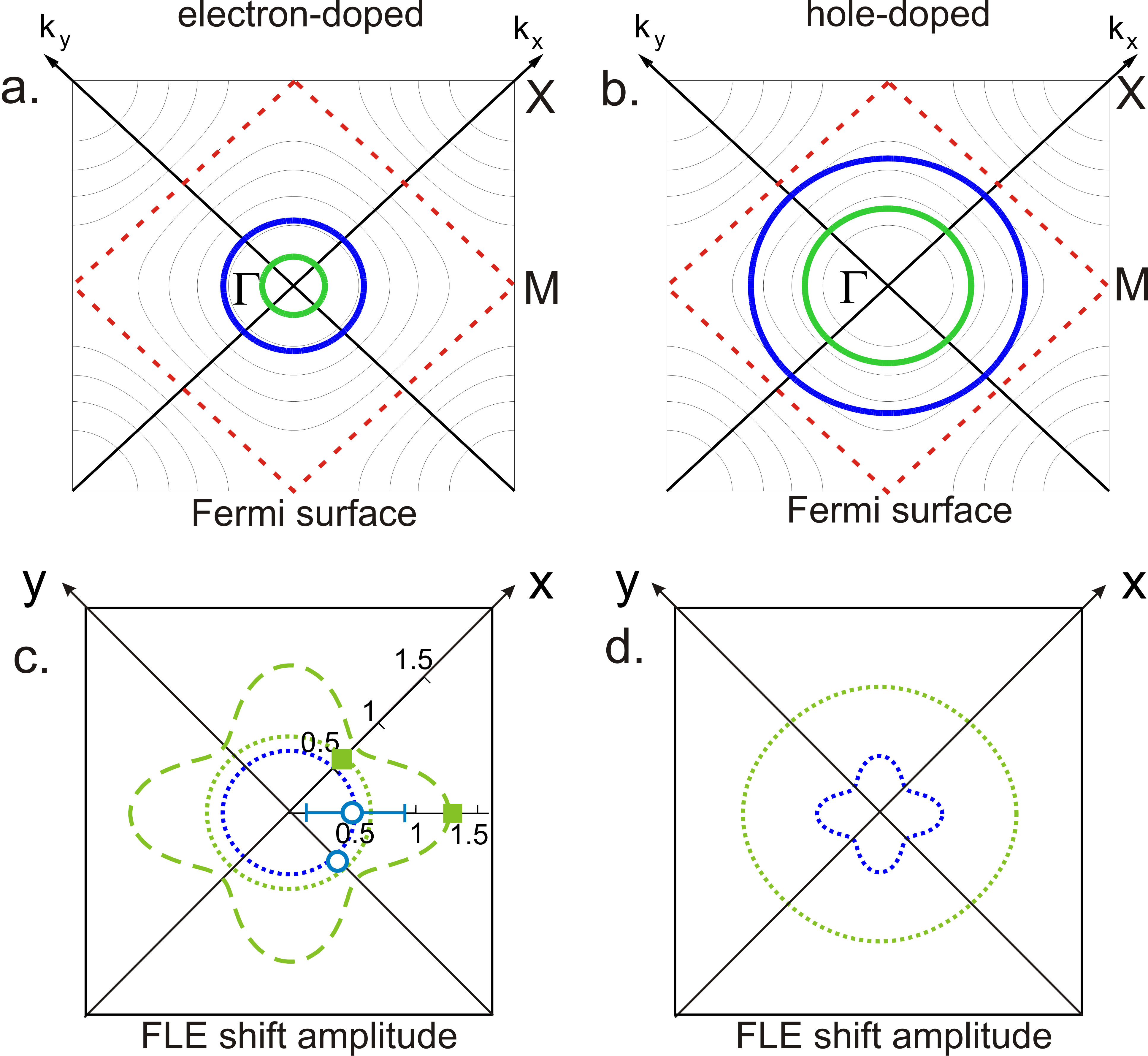} 
\caption{(a) Inner ($\alpha$, green) and outer ($\beta$, blue) hole-like pockets in the folded Brillouin zone of Ba(Fe$_{1-x}$Co$_x$)$_2$As$_2$; the black lines represent the order parameter contour lines corresponding to the $s\pm$ symmetry: the nodal lines (zero value) are the red dashed lines. (b) Example of a similar situation with larger Fermi surfaces (hole-doped compounds). (c) expected gap amplitude dispersion (in dotted lines) for the situation described in (a), and experimental points; green squares correspond to the $\alpha$ band and blue circles to the $\beta$ one. The green dashed line is a guide to the eye illustrating a behavior for $\alpha$ compatible with the experimental data (d) expected gap amplitude dispersions for the situation described in (b).} 
\label{theo}
\end{figure}

We preferred to carefully fit the EDCs with Fermi functions in order to precisely quantify the FLE shift - which is proportional to the gap - and measure this value for high symmetry directions in k-space. The fitting functions together with the experimental spectra for the $\alpha$ band along $\Gamma M$ are shown in Fig.~\ref{fitting} (a)-(b), and the difference between the two fitting functions across T$_c$ in Fig.~\ref{fitting} (c). The FLE shifts obtained by this procedure are the following: along $\Gamma M$, we obtain for $\beta$ 0.5$\pm$0.4 meV and for $\alpha$ 1.3$\pm$0.4 meV. On the other hand, along $\Gamma X$ the FLE shifts are 0.55$\pm$0.4 for $\beta$ and 0.6$\pm$0.4 for $\alpha$, as summarized in Fig.~\ref{theo} (c).

This small but unambiguous difference between the FLE shifts of $\alpha$ and 
$\beta$ deserves further attention, but first the question should be 
addressed if other factors than the opening of the SC gap can affect such shifts. 
One important factor that should be kept in mind is that the FLE is directly affected by the scattering rate: the stronger the scattering, the broader the superconducting coherent peak, and thus the smaller the Fermi leading edge. The scattering can be in general band- and momentum-dependent, and may produce an effective gap anisotropy along $\Gamma M$ and $\Gamma X$. 

We tried to evaluate these possible effects on our data by using a BCS-like
function to fit the differences in the EDC's presented in 
Fig.~\ref{gap_GM} (g) and (h), in order to model the transfer of spectral weight into the superconducting coherent peak for the two bands along $\Gamma M$. As already mentioned, the absence of a marked SC gap makes it hard to unambiguously fit the superconducting coherent peak. Once the hypothesis on the SC peak line shape is made (BCS in our case), fitting the SC-normal differences is instead a very sensitive way of obtaining information on the coherent peak position - the price 
to pay is of course the absence of any information on the overall density of states in the two phases. Our fits show that the differences for both $\alpha$ and $\beta$ can be correctly reproduced only by using the same values for the full width of the coherent peak (3.5 meV) in the two cases, but with different values for its position (3.5 meV and 2 meV for $\alpha$ and $\beta$, respectively). This suggests that the difference in scattering rate should not be regarded as the main factor in determining the inequivalent behavior for $\alpha$ and $\beta$ along $\Gamma M$; furthermore, it corroborates the hypothesis that the FLE shift is proportional to the SC gap, and indicates a slight anisotropy in the gap opening for the $\alpha$ band. 

Before discussing the possible implications of this small anisotropy on the superconducting order parameter, we present in Fig.~\ref{theo} (a) a schematic view of the hole-like pockets, together with the contour lines expected for an $s\pm$ order parameter with a $cos(k_x)+cos(k_y)$ behavior, and we show the corresponding k-dependence of the gap amplitude in  Fig.~\ref{theo} (c), in dotted lines. In Fig.~\ref{theo} (b) and (d) we present a similar situation with larger Fermi surfaces, which would be representative of a hole-doped compound. 


One should observe that the contour lines of the order parameter magnitude evolve from a circular shape around the $\Gamma$ point to a diamond shape (where its value is zero) for larger momenta. 
Therefore its magnitude on a small circular Fermi surface around $\Gamma$ (green curve in Fig.~\ref{theo} (a) and corresponding gap in Fig.~\ref{theo} (c)) will be nearly constant, while a flower-shaped modulation is expected for larger circular Fermi surfaces (blue curve in Fig.~\ref{theo} (b) and gap in Fig.~\ref{theo} (d)). Also the overall magnitude is expected to be smaller in the latter case, because the Fermi surface approaches the node line (red dashed contour lines). This scenario is indeed found for the hole-doped pnictides of the 122 family~\cite{Wray2008} (see Fig. \ref{theo}, (b) and (d)).

Conversely, our results do not fit into this simple picture: the experimental FLE shift values (blue circles in Fig.\ref{theo} (c)) for the larger Fermi surface ($\beta$, dark blue) are quite isotropic , while the experimental FLE shift values (green squares in Fig.\ref{theo} (c)) for the smaller Fermi surface ($\alpha$, in green in Fig.\ref{theo} (a)) are compatible with a flower-like modulation.


Our results are consistent with the symmetry expected within the $s\pm$ scenario~\cite{Mazin2008, Kuroki2008}. However, they also indicate that the simplest version of this model, in which the superconducting order parameter follows a pure cos($k_x$)$+$cos($k_y$) law, has to be extended~\cite{Kuroki2008}: in order to account for our findings, in particular for the modulation of the $\alpha$ band, a band dependent order parameter is needed. 

One possibility for this is to have higher harmonics contributing to the order parameter, with orbital dependent weighting prefactors. Higher harmonics are needed for the gap on the inner Fermi surface to modulate, because they have more node lines, and they are closer to $\Gamma$; these harmonics have however to contribute much less to the gap of the outer Fermi surface.
Such strong orbital dependence is in line with the analysis by Kuroki {\it et. al}~\cite{Kuroki2008}.

A few considerations should be made on the differences observed for  
the electron and hole doped compounds of the 122 family. First, as just discussed, the order parameter can be described with a single harmonic term for the hole doped systems, while multiple terms might be necessary for the electron doped ones. Then another important difference concerns the relation between gap amplitude and Fermi surface nesting:
in hole doped Ba$_{0.6}$K$_{0.4}$Fe$_2$As$_2$, the largest SC gap opens in the hole pocket presenting the most efficient nesting with electron pockets~\cite{Nakayama2011}, 
which can be easily explained by inter-band scattering favoring a larger SC order parameter amplitude.

In the case of electron-doped Ba(Fe$_{1-x}$Co$_x$)$_2$As$_2$, nesting with the electron pockets was found for the $\beta$ band~\cite{Brouet2009}, and this doesn't seem to be correlated to a larger gap; it may be related to a stronger scattering rate for $\beta$ than for $\alpha$, which would affect the effective FLE shift, but as discussed above this is unlikely to be the main factor
emerging from our experimental data. Furthermore, it should be kept in mind that the gap values presented here were measured for a $k_z$ different from the values explored in ref.~\cite{Brouet2009}. Since Ba(Fe$_{1-x}$Co$_x$)$_2$As$_2$
presents a marked three-dimensional character, an extensive $k_z$ dependent measurement of the gap (as performed in K-doped 122 compounds~\cite{Zhang2010, Xu2011}), would be needed to get clear indications on its relation with nesting. 


The differences between hole and electron doped compounds may come from the multi-orbital character of their electronic structure. While the inner hole pockets of K-doped 122's are formed by $d_{xz}$ and $d_{yz}$, the outer band is constituted of $d_{xy}$~\cite{Zhang2010}. On the other hand, the two inner bands of Co-doped 122's are constituted of $d_{xz}$ and $d_{yz}$, but the outermost one is hybridized in a more complex way, especially with $d_{z^2}$ away from $k_z\cong1~\left[4\pi/c\right]$~\cite{Mansart2011}. 
It should be noted that a recent ARPES study on both hole-doped and isovalent-doped 122 compounds by Shimojima \textit{et al.} shows evidence of an orbital independent order parameter amplitude for the different hole-like pockets~\cite{Shimojima2011}, while for the electron doped system presented here we find instead indications of orbital dependence. 
The importance of these considerations should trigger more studies on the symmetry of the order parameter comparing electron- and hole-doped compounds of the 122 family. For ARPES experiments, it would be particularly important to compare more extensive experimental results with detailed calculations of the electronic structure of these compounds: this would allow to unambiguously extract quantitative values for the SC gap, going beyond the level of uncertainty coming from the use of FLE shifts and taking into account the density of states of these material when modelling the transfer of spectral weight from the normal to the superconducting state.  

In conclusion, we performed an angle resolved photoemission study of the hole-like Fermi surface above and below the superconducting phase transition in optimally doped Ba(Fe$_{1-x}$Co$_x$)$_2$As$_2$ along high symmetry directions in k-space. The analysis of the Fermi leading edge shifts confirms that the SC gap is isotropic for the outer hole pocket $\beta$, while it shows slight anisotropies for the inner one, $\alpha$. These results are consistent with a $s\pm$ order parameter following a cos($k_x$)$+$cos($k_y$) behavior, and suggest that multiple harmonic terms may be included in its description. More comparative studies 
between hole and electron doped materials of the 122 family appear necessary to completely understand the implications of these results, and the dependence of the superconducting order parameter on the nature of the various orbitals contributing to the physics of these multiband materials. 
 
The authors gratefully acknowledge D. Lonza for his technical assistance during the experiments, and E. Cappelluti and M. Capone for interesting discussions.
The research leading to these results has received funding from the European Community's Seventh Framework Programme (FP7/2007-2013) under grant agreement nº 226716. L. de' Medici is financially supported by Agence Nationale de la Recherche under Program No. ANR-09-RPDOC-019-01 and by RTRA Triangle de la Physique.

\end{document}